\documentstyle[12pt]{article}

\newcommand{\EQ}{\begin{equation}}
\newcommand{\EN}{\end{equation}}

\newcommand{\ket}[1]{\left|#1\right\rangle}      % Ket-Zustand
\newcommand{\bear}{\begin{eqnarray}}
\newcommand{\ear}{\end{eqnarray}}

\begin{document}

\topmargin 0pt
\oddsidemargin 5mm
\newcommand{\NP}[1]{Nucl.\ Phys.\ {\bf #1}}
\newcommand{\PL}[1]{Phys.\ Lett.\ {\bf #1}}
\newcommand{\NC}[1]{Nuovo Cimento {\bf #1}}
\newcommand{\CMP}[1]{Comm.\ Math.\ Phys.\ {\bf #1}}
\newcommand{\PR}[1]{Phys.\ Rev.\ {\bf #1}}
\newcommand{\PRL}[1]{Phys.\ Rev.\ Lett.\ {\bf #1}}
\newcommand{\MPL}[1]{Mod.\ Phys.\ Lett.\ {\bf #1}}
\newcommand{\JETP}[1]{Sov.\ Phys.\ JETP {\bf #1}}
\newcommand{\TMP}[1]{Teor.\ Mat.\ Fiz.\ {\bf #1}}
     
\renewcommand{\thefootnote}{\fnsymbol{footnote}}
     
\newpage
\setcounter{page}{0}
\begin{titlepage}     
\begin{flushright}
ITFA-96-53
\end{flushright}
\vspace{0.5cm}
\begin{center}
\large{ Exact solution of the lattice vertex model analog of 
the coupled Bariev $XY$ chains.}\\
\vspace{1cm}
\vspace{1cm}
 {\large M. J.  Martins$^{1,2}$  and P.B. Ramos$^{2}$} \\
\vspace{1cm}
\centerline{\em ${}^{1}$ Instituut voor Theoretische Fysica, Universiteit van Amsterdam  }
\centerline{\em  Valcknierstraat 65, 1018 XE Amsterdam, The Netherlands}
\centerline{ \em and }
\centerline{\em ${}^{2}$ Departamento de F\'isica, Universidade Federal de S\~ao Carlos}
\centerline{\em Caixa Postal 676, 13565-905, S\~ao Carlos, Brasil}
\vspace{1.2cm}   
\end{center} 
\begin{abstract}
We present the algebraic Bethe Ansatz solution for the vertex model 
recently proposed by Zhou \cite{ZH} as the classical analog of the Bariev 
interacting XY chains. The relevant commutation rules between the
creation fields  
contain the Hecke symmetry pointed out 
recently by Hikami and Murakami \cite{HK}. The
eigenvalues of the corresponding transfer matrix are explicitly given.
\end{abstract}
\vspace{.2cm}
%\centerline{PACS numbers: 05.50+q, 64.60.Cn, 75.10.Hk, 75.10.Jm}
\vspace{.2cm}
\centerline{November 1996}
\end{titlepage}

\renewcommand{\thefootnote}{\arabic{footnote}}

Some years ago, Bariev has formulated a model of interacting $XY$ chains and 
solved it exactly, in one-dimension, by the coordinate Bethe Ansatz approach \cite{BA}.
The model couples two $XY$ models and its Hamiltonian on a lattice of length $L$
can be written as \cite{BA,BA1,ZH}
\EQ
H= \sum_{i=1}^{L} \left \{ ( \sigma_{i}^{+} \sigma_{i+1}^{-}  
+ \sigma_{i}^{-} \sigma_{i+1}^{+}  ) \exp(\alpha \tau_{i+1}^{+} \tau_{i+1}^{-} )
+ ( \tau_{i}^{+} \tau_{i+1}^{-}  
+ \tau_{i}^{-} \tau_{i+1}^{+}  ) \exp(\alpha \sigma_{i}^{+} \sigma_{i}^{-} )
\right \}
\EN
where $\sigma_{i}^{\pm}$ and $\tau_{i}^{\pm}$ are two commuting sets of Pauli
matrices acting on site $i$, and $\alpha $ is the coupling constant. Soon after
that, Bariev \cite{BA1} has generalized this model to include many 
coupled $XY$ chains. After a Jordan-Wigner transformation, the 
model (1) and its generalizations, 
can be seen as an electronic system where the hopping term depends
(asymmetrically) on the occupation number of the site itself \cite{BA2,BA3}. In this
sense, the coordinate Bethe Ansatz solution of 
these models has been used in refs. \cite{BA2,BA3}  
to explore the finite size behaviour, the excitations as well as some related
conductivity properties. 

The quantum integrability of (1), however, has only 
been recently proved by Zhou \cite{ZH}
in terms of the quantum inverse scattering 
approach. Zhou \cite{ZH} was able to
construct the two dimensional vertex model whose transfer matrix 
commutes with the Bariev Hamiltonian (1). The purpose of 
this letter is to show that such underlying vertex model 
can be diagonalized by the algebraic 
Bethe Ansatz \cite{FA,REV}. We recall that  this method is a powerful
mathematical technique,  which can provide us 
with information concerning 
the properties of the Bariev chain within a unified
perspective.
Our formulation 
is strongly inspired by our recent construction of
the Bethe states of the Hubbard model by means of the quantum inverse scattering approach
\cite{HUB}. We remark, however, that the structure of Zhou's $R$-matrix is a bit different
than that appearing in the Hubbard model \cite{SA,OLL}. Indeed, we shall see that the
appropriate parametrization of the quantum $R$-matrix  associated to
the  Bariev chain is different and, in fact, much simpler
than that we have found for the Hubbard model \cite{HUB}.

The quantum $R$-matrix $R(\lambda,\mu)$ 
found by Zhou \cite{ZH} has fifteen non-zero 
Boltzmann weights, and 
following ref. \cite{ZH} we denote them by $\rho_i(\lambda,\mu)$,
$i=1, \cdots, 15 $. However, many of the weights are related 
to each other under some functional properties, such as 
$ \rho_{i}(\lambda,\mu)=-\rho_{j}(\mu, \lambda); 
\rho_{i}(\lambda,\mu)= \rho_{j}(h \lambda,h \mu); 
\rho_{i}(\lambda,\mu)=
h \rho_{j}(\lambda,\mu )$  
\footnote{ For instance, from ref. \cite{ZH}, it is possible to check that 
$\rho_{11}(\lambda,\mu)=\rho_{9}( \lambda/h,\mu/h)$; 
$\rho_{4}(\lambda,\mu)= \rho_{2}(\lambda, \mu) \rho_{2}(\lambda/h, \mu/h)$; 
$\rho_{5}(\lambda,\mu)=
- \rho_{12}(\mu,\lambda)$.}. The parameter  $h$ is given in terms of 
the coupling constant $\alpha$
by $h =\exp(\alpha)$. For explicit expressions 
we refer to ref. \cite{ZH}. Here we quote
only few extra identities which we found relevant in 
the course of our calculations. These
are given by
\EQ
\rho_{15}(\lambda,\mu)[ \rho_{9}(\lambda,\mu)+ \rho_1(\lambda,\mu) ] = 
\rho_{5}(\lambda,\mu)
\rho_{6}(\lambda,\mu),~~
\rho_{6}(\lambda,\mu) \rho_{1}(\lambda,\mu)+ \rho_{5}(\lambda,\mu)  
\rho_{15}(\lambda,\mu) = \rho_{6}(\lambda,\mu)
\rho_{7}(\lambda,\mu)
\EN
\EQ
\rho_{12}(\lambda,\mu)[ \rho_{9}(\lambda,\mu)+ \rho_{1}(\lambda,\mu) ] = \rho_{5}(\lambda,\mu)
\rho_{4}(\lambda,\mu),~~
\rho_{5}(\lambda,\mu) \rho_1(\lambda,\mu)+ \rho_{15}(\lambda,\mu)  
\rho_{6}(\lambda,\mu) = \rho_{5}(\lambda,\mu)
\rho_{10}(\lambda,\mu)
\EN

In order to diagonalize the transfer-matrix of the classical 
vertex model corresponding to the  Bariev chain 
we can basically follow the main steps
of our recent algebraic construction of the 
Bethe Ansatz for the Hubbard model \cite{HUB}.
We take as the reference state $\ket{0}$, 
the standard ferromagnetic vacuum where all
the spins  are in the ``up'' eigenstate of $\sigma_j^z$ and $\tau_j^z$. 
We solve the 
Yang-Baxter algebra for  Zhou's $R$-matrix 
by writing the monodromy matrix $\cal{T}(\lambda)$
in the auxiliary space as 
\EQ
{\cal T}(\lambda) =
\pmatrix{
B(\lambda)       &   \vec{B}(\lambda)   &   F(\lambda)   \cr
\vec{C}(\lambda)  &  \hat{A}(\lambda)   &  \vec{B^{*}}(\lambda)   \cr
C(\lambda)  & \vec{C^{*}}(\lambda)  &  D(\lambda)  \cr}
\EN
where $\vec{B}(\lambda)$ $(\vec{B^{*}}(\lambda))$ and 
$\vec{C}(\lambda)$ $(\vec{C^{*}}(\lambda))$  are two component vectors 
with dimensions $1 \times 2$$(2 \times 1)$ and $2 \times 1$$(1 \times 2)$, respectively. The 
operator $\hat{A}(\lambda)$ is a $2 \times 2$ 
matrix and the other remaining operators are scalars. The transfer matrix $T(\lambda)$ is
the trace of $\cal{T}(\lambda)$ on the auxiliary space, and the eigenvalue problem 
becomes
\EQ
[B(\lambda)+\sum_{a=1}^{2}A_{aa}(\lambda)+D(\lambda)]\ket{ {\Phi}_{n}(\lambda_{1},\cdots ,\lambda_{n})} =
\Lambda(\lambda,\{\lambda_{i}\}) \ket{ {\Phi}_{n}(\lambda_{1},\cdots ,\lambda_{n})}
\EN

The set of variables $\{ \lambda_1, \cdots, \lambda_n \} $ parametrizes the
multi-particle Hilbert space by the action of the creation fields on the reference
state ${\ket{0}}$. The operators $\vec{B}(\lambda)$, 
$\vec{B^{*}}(\lambda)$ and $F(\lambda)$ play 
the role of creation fields 
while $\vec{C}(\lambda)$,$\vec{C^{*}}(\lambda)$, 
$C(\lambda)$ and 
$A_{ab}(\lambda)$, for $a \neq b=1,2 $, are annihilators.  This
means that the monodromy matrix (4) has a triangular form when acting on the
reference state ${\ket{0}}$. In addition, we have the following ``diagonal''
identities
\EQ
B(\lambda)\ket{0} = \ket{0},~~ D(\lambda)\ket{0} = 
{[\lambda]}^{2L}\ket{0},~~ 
A_{aa}(\lambda)\ket{0} =[\lambda h]^{L} \ket{0} , a=1,2
\EN

A crucial step in 
algebraically solving the eigenvalue problem (5) is to find the
appropriate commutation rules between two fields of $\vec{B}(\lambda)$ or 
$\vec{B^{*}}(\lambda)$ type. Similar to what  happens 
for the Hubbard model \cite{HUB}, their 
commutation rules are equivalent, because they generate as a new 
operator only the
common creation field $F(\lambda)$. Remarkably enough, these 
commutation rules already encode
the basic underlying hidden symmetry of the Bariev chain \cite{HK}.
We can see this, for instance, in the 
commutation relation \footnote{ We remark that identities
such as (2) and (3) are important in the simplification of the commutation rules.}
between the fields $\vec{B}(\lambda)$ and $\vec{B}(\mu)$
\EQ
\vec{B}(\lambda) \otimes \vec{B}(\mu) = 
\vec{B}(\mu) \otimes \vec{B}(\lambda) \hat{r}(\lambda,\mu)
%\nonumber \\
+ \frac{\vec \xi}{\rho_{9}(\lambda,\mu) } 
\{ \rho_{5}(\lambda,\mu ) F(\lambda)B(\mu) + \rho_{5}(\mu,\lambda)F(\mu)B(\lambda) \} 
\EN
where 
the vector 
$\vec{\xi}$ and the matrix $\hat{r}(\lambda,\mu)$ have the following structures
\EQ
{\vec \xi} = 
\matrix{(
0  &1  &h^{-1}  &0 )  \cr},~
\hat{r}(\lambda,\mu) = 
\pmatrix{
1  &0  &0  &0  \cr
0  &a(\lambda,\mu)  &b(\lambda,\mu)  &0  \cr
0  &b(\lambda,\mu)  &\tilde{a}(\lambda,\mu)  &0  \cr
0  &0  &0  &1  \cr}
\EN
and functions $a(\lambda,\mu)$, $\tilde{a}(\lambda,\mu)$ and $b(\lambda,\mu)$ are
given by
\EQ
a(\lambda,\mu) = \frac{\lambda (1-h^2)}{\lambda -h^2 \mu}
,~~ \tilde{a}(\lambda,\mu)= \frac{\mu (1-h^2)}{\lambda -h^2 \mu},~~ 
b(\lambda,\mu) = -\frac{h (\lambda-\mu)}{\lambda-h^2 \mu}
\EN

The structure of the Boltzmann weights of the matrix $\hat{r}(\lambda,\mu)$ 
are the same of that appearing in the $6$-vertex model with an azimuthal anisotropy $\eta$
given by $\eta=i\ln(h)=i\alpha$. In order to see that, we have to introduce the following 
parametrization
\EQ
\lambda = \exp[i k(\lambda)]
\EN
and consequently the Boltzmann weights can be rewritten in terms of the 
difference $k= k(\lambda) -k(\mu)$ and the anisotropic constant $\eta$ as
\EQ
a(k)= \frac{\exp(ik/2) \sin(\eta)}{\sin(k/2+\eta)},~~
\tilde{a}(k)= \frac{\exp(-ik/2) \sin(\eta)}{\sin(k/2+\eta)},~~
b(k)= -\frac{ \sin(k/2)}{\sin(k/2+\eta)}
\EN

By means of a transformation which preserves the Yang-Baxter equation, the so-called symmetry
breaking transformation \cite{WA}, the Boltzmann weights $a(k)$ and $\tilde{a}(k)$ can
be symmetrized in order to give the standard anisotropic $6$-vertex model. In fact, the
presence of the asymmetric version of the $6$-vertex model in the commutation rules is
a clear sign that the underlying symmetry is of Hecke type. It is not
difficult to see, from the asymmetric 
vertex model (8), that we can construct a braid operator which   appears as the generator 
of the Hecke algebra \cite{WA}. We remark here that such symmetry was first noticed by
Hikami and Murakami \cite{HK} 
in the context of the lattice Schr\"odinger equation for the ``fermionic''
formulation of the Bariev 
Hamiltonian (1). The only subtle point is 
the minus sign on weight $b(k)$. The Yang-Baxter
equation is invariant under $b(k) \rightarrow -b(k) $, and 
the sign $\pm$ can be
interpreted as periodic/antiperiodic boundary 
conditions when the size of the quantum 
Hilbert space is odd \cite{HUB1} \footnote{ For an even size the sign does not matter. We also recall that
in the fermionic formulation of the Bariev chain this sign is positive.}.

Now, the construction of the eigenvectors and the eigenvalues 
goes fairly parallel to the
formulation we have recently presented for the Hubbard model \cite{HUB}. Basically, we have
to adapt the `` diagonal '' commutation rules of ref. \cite{HUB} 
by taking into account the specific weights of Zhou's 
$R$-matrix and also  consider the
convenient parametrization given in equation (10).  Here we only 
present 
our final results for  
the eigenvalues of the ``covering''
vertex model. Many other results, as well as the main 
technical steps we have developed will
be presented in a separate publication \cite{HUB1}, together 
with the detailed algebraic
solution of the Hubbard model \cite{HUB}. We remark that, recently,  
the exact expression for the eigenvalue appears to be very important in the
study of finite 
temperature properties of integrable models \cite{KO,KU,DV}. We found
that the eigenvalue of the transfer matrix associated to  the Bariev chain is
given by
\EQ
\Lambda(\lambda,\{\lambda_{i}\}) = 
\prod_{i=1}^{n} \frac{h^{-1} +h \lambda_i \lambda}{\lambda_i -\lambda} + 
{\lambda}^{2L}
\prod_{i=1}^{n} \frac{1+h^2 \lambda_i \lambda}{\lambda-h^2 \lambda_i} +  
%\nonumber \\
[\lambda h]^{L} \prod_{i=1}^{n} \frac{h^{-1} +h \lambda_i \lambda}{\lambda- \lambda_i} 
\Lambda^{(1)}(\lambda,\{\lambda_{i}\}) 
\EN
where $\Lambda^{(1)}(\lambda,\{\lambda_{i}\})$ is the eigenvalue of the vertex model
defined by the auxiliary $R$-matrix $\hat{r}(\lambda,\mu)$ in the presence of
inhomogeneities. Furthermore the variables $\{ \lambda_i \} $ are constrained by the Bethe
Ansatz equation
\EQ
[\lambda_i h]^{-L}= - (-1)^{n}
\Lambda^{(1)}(\lambda=\lambda_i,\{\lambda_{j}\}) ,~~ i=1, \cdots, n
\EN

The auxiliary  problem can be solved 
by using the standard $6$-vertex formulation
of Faddeev et al \cite{FA,REV}, adapted to include the inhomogeneities $\{ \lambda_i \} $.
In the diagonalization procedure, it is necessary to introduce the auxiliary variables 
$ \{ \mu_j \} $ and
the eigenvalue 
$\Lambda^{(1)}(\lambda,\{\lambda_{i}\},\{ \mu_j \}) $ reads
\EQ
\Lambda^{(1)}(\lambda,\{\lambda_{i}\},\{ \mu_j \}) =
\prod_{j=1}^{m} \frac{1}{b(\mu_j,\lambda)} +
\prod_{i=1}^{n} b(\lambda,\lambda_i) \prod_{j=1}^{m} \frac{1}{b(\lambda,\mu_j)}
\EN
where the variables $ \{ \mu_j \} $ satisfy the equation
\EQ
\prod_{i=1}^{n} b(\mu_j,\lambda_i) = - \prod_{k=1}^{m} \frac{b(\mu_j,\mu_k)}{b(\mu_k,\mu_j)},~~
j=1, \cdots, m
\EN

Finally, all these results can be combined in order to give us the eigenvalue and
the Bethe Ansatz equations. At this point, to cast the final results in a convenient form,
we redefine the variables $\lambda$, $ \{ \lambda_i \} $, and $\{ \mu_j \}$ by
\EQ
\lambda_i h = \exp(ik_i), \mu_j = \exp( i \Lambda_j), \lambda= \exp(ik)
\EN

In terms of these new parameters the expression for the eigenvalue is
\bear
\Lambda(k,\{k_{i}\},\{ \Lambda_j \} ) = \prod_{i=1}^{n} \frac{ \cos(k/2 +k_i/2 -\eta/2)}
{i \sin( k_i/2 -k/2 +\eta/2)} 
+ \exp{(i2L k)} \prod_{i=1}^{n} \frac{\cos( k_i/2 +k/2 -\eta/2)}{i \sin(k/2 -k_i/2 +\eta/2)}
\nonumber \\
+ \exp{[i(k-\eta)L]} \left \{ \prod_{i=1}^{n} \frac{i \cos(k/2 +k_i/2 -\eta/2)}{
\sin(k_i/2 -k/2 +\eta/2)} \prod_{j=1}^{m} -\frac{ \sin(\Lambda_j/2 -k/2 +\eta)}{\sin(\Lambda_j/2
-k/2)} + 
 \right. \nonumber \\ \left.
\prod_{i=1}^{n} \frac{i \cos(k/2 +k_i/2 -\eta/2)}{
\sin(k/2 -k_i/2 +\eta/2)} \prod_{j=1}^{m} 
-\frac{ \sin(k/2 - \Lambda_j/2  +\eta)}{\sin(k/2 - \Lambda_j/2)}
\right \}
\ear
and the nested Bethe Ansatz equations are given by
\EQ
\exp(ik_i L) = - (-1)^{n-m} \prod_{j=1}^{m} \frac{ \sin(k_i/2 - \Lambda_j/2 +\eta/2)}
{ \sin(k_i/2 - \Lambda_j/2 -\eta/2)},~~ i=1, \cdots, n 
\EN
\EQ
(-1)^{n} \prod_{i=1}^{n} \frac{ \sin(\Lambda_j/2 -k_i/2 -\eta/2)}
{\sin(\Lambda_j/2 -k_i/2 +\eta/2)} = - \prod_{k=1}^{m} \frac{\sin(\Lambda_j/2 -\Lambda_k/2- \eta)}
{\sin(\Lambda_j/2 -\Lambda_k/2+ \eta)},~~ j=1, \cdots, m
\EN

This last equation is similar to that found early by Bariev \cite{BA,BA1}  in the context of
the coordinate Bethe Ansatz approach,  as should be. In order to recover the results of
Bariev \cite{BA,BA1} for the eigenenergies 
$E(L)$ of the Hamiltonian (1), we just have to take
the logarithmic derivative of the transfer matrix 
eigenvalue at the point $\lambda = 0$. This 
calculation leads us  to
\EQ
E(L) = 2h \sum_{i=1}^{L} \cos(k_i)
\EN

We conclude this letter with the following remarks. The extra signs we have found in the
Bethe Ansatz equations (18,19) are typical of the `` bosonic '' formulation (1) of the
Bariev chain. They can be related with peculiar
boundary conditions \cite{HUB,RD}, and
they are not present if one formulates  the diagonalization problem for the
``fermionic '' version of (1). Our algebraic formulation 
has an invariance under 
$h \rightarrow h^{-1}$, which is in accordance with the symmetry of the Bariev chain (
$\alpha \rightarrow -\alpha$) \cite{BA}. Following the results of Shiroischi
and Wadati \cite{SWA}, there exists a way of generating a generalized
Bariev chain from Zhou's $R$-matrix. Defining the vertex 
operator \cite{SWA} ${\cal{L}}^{\theta_0}(\lambda) =
P R(\lambda,\theta_0)$, where
P is permutator and $R(\lambda,\theta_0)$ is 
Zhou's $R$-matrix, we can define
a one-parameter ($\theta_0$) family of vertex models by the transfer matrix
$T^{\theta_0}(\lambda)= Tr_a[ {\cal{L}}^{\theta_0}_{aL}(\lambda) 
\cdots {\cal{L}}^{\theta_0}_{a1}(\lambda) ] $. Such vertex model 
can be diagonalized following the basic steps we presented so far. The main
change is concerned with the action of the `` diagonal '' 
operators on the reference
state ${\ket{0}}$. In this case we find $B(\lambda) {\ket{0}} = [\rho_{1}(
\lambda,\theta_0)]^{L} {\ket{0}}$,
$ A_{aa}(\lambda){\ket{0}}= [\rho_{3}(\lambda,\theta_0)]^{L} {\ket{0}}  $,
$ D(\lambda){\ket{0}}= [\rho_{9}(\lambda,\theta_0)]^{L} {\ket{0}}  $, in such way that
only the terms which are proportional to the power of $L$ change in the
expressions (12) and (13). 
Finally, since the parametrization (10) is quite simple,
it seems interesting to re-investigate the Yangian symmetry  as well as
the analytical properties of the transfer matrix associated to the Bariev chain in
light of the recent results of refs. \cite{SF,YL}.

\section*{Acknowledgements}
The authors thank J. de Gier for helpfull comments. 
This  
work was support by  FOM (Fundamental Onderzoek der Materie) 
and Fapesp ( Funda\c c\~ao
de Amparo \'a Pesquisa do Estado de S. Paulo).

\end{document}